\documentclass[preprint,pre,eqsecnum,showpacs]{revtex4}

\usepackage{graphicx}

\begin{document}

\title[Stiff polymers on PF fractal families]
{Stiffness dependence of critical exponents of semiflexible  polymer  chains
 situated on two-dimensional compact fractals}

\author{Ivan \surname{\v Zivi\' c}}
\email{ivanz@kg.ac.rs}
\affiliation{
Faculty of Natural Sciences and Mathematics,
University of Kragujevac, 34000 Kragujevac, Serbia}
\author{Sun\v cica \surname{Elezovi\' c-Had\v zi\' c}}
\email{suki@ff.bg.ac.rs}
\affiliation{Faculty of Physics,
University of Belgrade, P.O.Box 44, 11001 Belgrade, Serbia}
\author{Sava \surname{Milo\v sevi\'c}}
\email{emilosev@etf.rs}
\affiliation{Faculty of Physics,
University of Belgrade, P.O.Box 44, 11001 Belgrade, Serbia}

\date{\today}

\begin{abstract}
We present an exact and Monte Carlo
renormalization group (MCRG) study of semiflexible polymer chains on an infinite
family of the plane-filling (PF) fractals. The fractals are compact, that is, their fractal dimension $d_f$ is
equal to 2 for all members of the fractal family enumerated
by the odd integer $b$ ($3\le b< \infty$). For various values of stiffness parameter $s$ of the chain, on the PF fractals (for $3\le b\le 9$) we calculate exactly
the critical exponents $\nu$ (associated with the mean squared end-to-end distances of polymer chain) and $\gamma$ (associated with the total number of different polymer chains). In addition, we calculate $\nu$ and $\gamma$  through the MCRG approach for $b$ up to 201. Our results show that, for each particular $b$, critical exponents are stiffness dependent functions, in such a way that the stiffer  polymer chains (with smaller values of  $s$) display enlarged  values of $\nu$, and diminished   values of $\gamma$. On the other hand, for any specific $s$,
the critical exponent $\nu$ monotonically  decreases, whereas  the critical exponent $\gamma$ monotonically increases, with the scaling parameter $b$.
We reflect on a possible relevance of the  criticality of  semiflexible polymer chains  on the PF family of fractals to the same problem on the
regular  Euclidean lattices.
\end{abstract}

\maketitle

\section{Introduction}
\label{uvod}

The self--avoiding walk (SAW)  is a random walk that must not contain self--intersections. It has been extensively studied as a challenging problem in statistical physics, and, in particular, as a satisfactory model of a linear   polymer chain \cite{vc}. The pure SAW is a good model for perfectly flexible polymer, where we ignore the apparent rigidity of real polymer, and, consequently,  to each  step of SAW we associate the same weight factor (fugacity) $x$. In most real cases, the  polymers are semiflexible with the various degree of stiffness. To take into account this property of polymers, in the continuous space models the stiffness of the SAW path is modeled by constraining the angle between the consecutive bonds of polymer, while in the lattice models, an energy barrier for each bend of the SAW is introduced. The lattice semiflexible SAW model (also known as persistent or biased SAW model), has been studied some time ago in a series of papers  \cite{halley}, with a focus on the  so-called rod-to-coil crossover. Afterwards, it was modified in various ways, in order to describe relevant aspects of different phenomena, such as protein folding  \cite{DoniachGarelOrland,bastola}, adsorption of semiflexible homopolymers \cite{kumar4}, transition between the disordered globule and the crystalline polymer phase \cite{n3,prellberg}, behavior of semiflexible polymers in confined spaces \cite{n2,n4}, or influence of an external force on polymer systems  \cite{kumar2,kumar1,kumar3,lam}.

In spite of numerous studies, a scanty collection of exact results
for semiflexible  polymers has been achieved so far, even for the
simplest lattice models. A few cases in which some properties of
semiflexible SAW can be studied exactly are: directed semiflexible
SAWs on regular lattices \cite{privmanKnjiga,kumar4}, and
semiflexible SAWs (with no constraints on the direction) on some
fractal lattices \cite{giacometti,tuthill}. In particular, exact
values of the end-to-end critical exponent $\nu$ and the entropic
exponent $\gamma$ were obtained for these models, and it turned
out that  in some cases critical exponents are universal, whereas
in other cases they depend on the stiffness of the polymer chain.
Universality arguments, as well as results of approximate and
extrapolation methods for similar models suggest that critical
exponents on regular (Euclidean) lattices should not be affected
by the value of the polymer stiffness. On the other hand, it is
not known what are the effects of rigidity on the critical
behavior of SAWs in nonhomogeneous environment. In order to
explore further this issue, in this paper we perform the relevant
study on the infinite family of the plane-filling (PF) fractal
lattices \cite{d1,zivic3}, which allow for an exact treatment of
the problem. These fractals appear to be compact, that is, their
fractal dimension $d_f$ is equal to 2. Members of the family can
be enumerated by an odd integer $b$ ($3\le b<\infty$), and as
$b\to\infty$ characteristics of these fractals approach, via the
so-called fractal--to-Euclidean crossover \cite{d2,d3}, properties
of the regular 2D lattice. By applying the exact real-space
renormalization group (RG) method \cite{ex1,d4}, as well as Monte
Carlo renormalization group (MCRG) method \cite{mc1,mc2,redner,d6}, we
calculate critical exponents $\nu$ and $\gamma$. We have performed
our calculations for as many as possible members of the fractal
family, for various degree of polymer stiffness, in order to study
consequent stiffness dependence of the critical exponents, as well
as to see the asymptotic  behavior of the exponents  in the
fractal--to--Euclidean crossover region.

This paper is organized as follows. We define the PF
family of fractals in Sec.~\ref{druga}, where we also present
the framework of our exact and MCRG approach to the evaluation
of the critical exponents $\nu$ and $\gamma$  of stiff polymers on
the PF fractals, together with the specific results.  In
Sec.~\ref{treca} we analyze the obtained data for the critical
exponents, and present an overall discussion and pertinent conclusions.

\section{Semiflexible polymers on the plane-filling fractal lattices}
\label{druga}

\begin{figure}
\centerline{\includegraphics[width=3.3in]{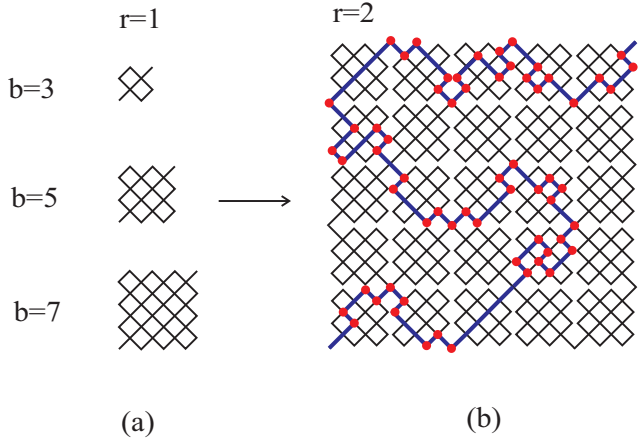}}
\caption{(Color online) (a) The first three fractal generators ($r=1$) of the
plane-filling (PF) family of fractals.
(b) The fractal structure of the $b=5$
PF fractal at the second stage of construction,
with an example of a piece of a possible SAW path
(thick line). The full dots represent  the  turn points of the  walker (that is,
  the bends of the SAW path), to which we associate the
 Boltzmann factor $s=e^{-\Delta/k_BT}$, where
$\Delta>0$ is the energy of SAW bend. Thus, for example, the presented SAW
configuration should contribute the weight $x^{97}s^{62}$  in the
corresponding RG equation (more specifically, in equation
(\ref{eq:RGB}) for $r=0$).}
 \label{fig1}
\end{figure}

 In this section we are going to apply the exact RG and the
MCRG method to calculate asymptotic properties of semiflexible
polymer chains  on the PF fractal lattices. Each member of the PF
fractal family is labelled by an odd integer $b$ ($3\le b
<\infty$), and can be constructed in stages. At the initial stage
($r=1$) the lattices are represented by the corresponding
generators (see Fig.~\ref{fig1}). The $r$th stage fractal
structure can be obtained iteratively in a self-similar way, that
is, by enlarging the generator by a factor $b^{r-1}$ and by
replacing each of its segments with the $(r-1)$th stage structure,
so that the complete fractal is obtained in the limit $r \to
\infty$. The shape of the fractal generators and the way the
fractals are constructed imply that each member of the family has
the fractal dimension $d_f$ equal to 2. Thus, the PF fractals
appear to be compact objects embedded in the two-dimensional
Euclidean space, that is, they resemble square lattices with
various degrees of inhomogeneity distributed self-similarly.

In order to describe stiffness of the polymer chain, we introduce the  Boltzmann factor
 $s={\mathrm{e}}^{-\Delta /k_BT}$, where
$\Delta$ is an  energy barrier associated with each bend of the SAW path, and $k_B$ is the Boltzmann constant.
 For   $0<s<1$ $(0<\Delta<\infty)$ we deal with  the   semiflexible polymer chain, whereas in the limits   $s=1$ $(\Delta=0)$ and $s=0$ $(\Delta=\infty)$ the polymer is a flexible chain or a rigid  rod, respectively.
 If we assign the weight $x$ to each step of the SAW, then
the weight of a walk having $N$ steps, with $N_b$ bends, is $x^Ns^{N_b}$, and consequently, the general form of the  SAW generating function can be written as
\begin{equation}\label{gen}
 G(x,s)=\sum_{N=1}^\infty \sum_{N_b=0}^{N-1} C(N,N_b) x^N
 s^{N_b}\>,
\end{equation}
where $C(N,N_b)$ is the number of $N$--step SAWs having $N_b$
bends.   For large $N$, it is generally expected \cite{vc} that
the total number $C(N,s)=\sum_{N_b=0}^{N-1} C(N,N_b)
 s^{N_b}$ of $N$--step SAW  displays  the following power law
\begin{equation}\label{g-def}
 C(N,s) \sim \mu ^N N^{\gamma -1}\>,
\end{equation}
where $\gamma$ is the entropic critical exponent, and $\mu$ is the connectivity constant.
Accordingly, at the critical fugacity $x_c=1/\mu(s)$, we expect the following singular behavior of the above generating function
\begin{equation}\label{g-sing}
  G_{sing}\sim
{(x_c-x)}^{-\gamma}\>.
\end{equation}

\begin{figure}
\centerline{\includegraphics[width=3.3in]{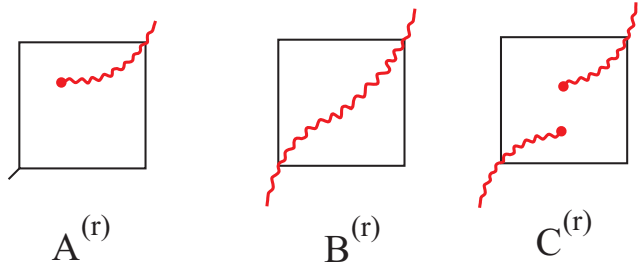}}
\caption{ \label{fig2} (Color online)  A diagrammatic representation of the
three restricted partition functions for an $r$th order of
the fractal construction of a member of the PF family.
The fractal interior structure is not shown. Thus, for
example, $A^{(r)}$ represents the SAW path that starts
somewhere within the fractal structure and leaves it at its
upper right link to rest of fractal.}
\end{figure}

On the other hand,  due to the self-similarity of the underlying structure,
an arbitrary SAW configuration on the PF fractals can be described, by
using the three restricted generating functions  $A^{(r)}$, $B^{(r)}$ and
$C^{(r)}$ (see Fig.~\ref{fig2}),
which represent partial sums of
statistical weights  of  all  feasible  walks within  the
$r$th stage fractal structure for the three possible kinds of SAWs.
One may verify that, for arbitrary $b$, the generating
function $G(x,s)$ is of the form
\begin{eqnarray}
G(x,s)=\sum_{r=1}^\infty {1\over b^{2r}}&\Biggl\{& \!\!g_1(B^{(r-1)},s)
\left[ A^{(r-1)}\right] ^2 \nonumber\\&+& \!\!g_2(B^{(r-1)},s)
\left[ C^{(r-1)}\right] \Biggr\}\>,\label{generatrisa}
\end{eqnarray}
where the coefficients $g_1(B^{(r-1)},s)$ and $
g_2(B^{(r-1)},s)$ are polynomials in $B^{(r-1)}$ and $s$.
This structure  for $G(x,s)$ stems from the fact that all
possible open SAW paths  can be made in only two ways,
using the $r$th order structures.
The  functions $A^{(r)}$, $B^{(r)}$ and
$C^{(r)}$ appear to be  parameters in
the corresponding recursion (renormalization group)
equations, which  have the form
\begin{eqnarray}
 A^{(r+1)}&=&a(B^{(r)},s) A^{(r)}\> ,
\label{eq:RGA}\\
 B^{(r+1)}&=&b(B^{(r)},s)\>  ,
\label{eq:RGB}\\
 C^{(r+1)}&=&c_1(B^{(r)},s){(A^{(r)})}^2+c_2(B^{(r)},s)C^{(r)},\label{eq:RGC}
\end{eqnarray}
where  the  coefficients
$a(B^{(r)},s)$, $b(B^{(r)},s)$ and $c_i(B^{(r)},s)$ $(i=1,2)$, are
polynomials in terms of $B^{(r)}$ and $s$, and do not depend on
$r$. The established RG transformation should be supplemented with the
initial conditions:
$ A^{(0)} = \sqrt{x}\>,$ $B^{(0)} = x$,   and $C^{(0)} =0$,
that are pertinent to the fractal unit segment.

The basic asymptotic properties of  SAWs are characterized by two
critical exponents $\nu$ and $\gamma$. The critical exponent $\nu$
is associated with the scaling law $\left <R_N^2\right>\sim
N^{2\nu}$ for the mean squared end-to-end distance  for $N$-step
SAW, while the critical exponent $\gamma$ is associated with the
total number of distinct SAWs described by (\ref{g-def}), for very
large $N$ \cite{vc}. We start by applying the above RG approach
to find the critical exponent $\nu$ for semiflexible polymers on
PF fractals. We shall first present the corresponding exact
calculation, and then we shall  expound on the MCRG approach. To
this end, we need to analyze (\ref{eq:RGB}) at the corresponding
fixed point. It can be shown that $b(B^{(r)},s)$ in (\ref{eq:RGB})
is a
 polynomial
\begin{equation}\label{RGB}
B'=\sum_{N,N_b}P(N,N_b) B^{N}s^{N_b}\>  ,
\end{equation}
where we have used the prime for the  $r$th order partition function
and no indices for the $(r-1)$th order partition function.
 Knowing the RG equation (\ref{RGB}), value of
the critical exponent $\nu$ follows from the formula \cite{d4}
\begin{equation}\label{ni}
\nu={\ln b\over \ln \lambda_\nu}\> ,
\end{equation}
where $\lambda _\nu$ is the relevant eigenvalue of the RG
equation (\ref{RGB}) at the nontrivial fixed point $0<B^*(s)<1$, that is
\begin{equation}\label{lamdai}
\lambda_\nu={ dB'\over dB}\biggm|_{B^*}\> .
\end{equation}
Consequently, evaluation of $\nu$ starts with determining the
coefficients $P(N,N_b)$ of (\ref{RGB}) and finding the pertinent
fixed point value $B^*(s)$, which is, according to the initial
condition $B^{(0)}=x$, equal to the critical fugacity $x_c=1/\mu$.

We have been able to find exact
values of $P(N,N_b)$ for $3 \le b \le 9$. For the first two fractals ($b=3$ and 5) the RG equation (\ref{RGB}) has the form
\begin{eqnarray}
B'&=&B^3+2B^5s^4\>,\label{RGB3}\\
B'&=&B^5+12B^7s^4+
2B^9s^4+12B^9s^6+6B^9s^8+ \nonumber\\
&&4B^{11}s^6+8B^{11}s^8+2B^{13}s^8+
4B^{13}s^{10}+2B^{15}s^{12},\nonumber\\
 &&\label{RGB5}
 \end{eqnarray}
respectively, while for $b=7$ and 9 they are disposed within the Electronic Physics Auxiliary Publication Service (EPAPS) \cite{supp}.  Knowing $P(N,N_b)$, for a given $b$, we use
Eqs. (\ref{RGB})--(\ref{lamdai}) to learn $B^*(s)$ and the critical exponent $\nu(s)$.
For  the $b=3$ PF fractal  the critical fugacity $B^*$ and the critical exponent $\nu$  can be obtained in the closed forms as functions of the stiffness parameter
\begin{eqnarray}
 B^*_{b=3}(s)&=&{\sqrt{\sqrt{1+8s^4}-1}\over 2s^2}\>,\label{ng3a}\\
\nu_{b=3}(s)&=&{\ln 3\over\ln\left(5-{\sqrt{1+8s^4}-1\over 2s^4}\right)}\>,\label{ng3b}
\end{eqnarray}
while for  $b=5$, 7 and  9   they can only be calculated numerically. We have chosen the set of six values for the polymer stiffness parameter ($s=1, 0.9, 0.7, 0.5, 0.3$ and $s=0.1$) and the obtained exact values are presented in the Tables~\ref{tabela1} and \ref{tabela2} (together with the results obtained by the MCRG method).
\begin{table*}
\caption{ The exact ($3\le b\le 9$) and
MCRG ($11\le b\le 201$) results for the critical fugacities $B^*$  of the PF family of fractals.  Each MCRG entry of the Table has been
obtained by performing at least $10^5$ MC simulations.  The numbers in the brackets represent the MCRG errors concerning  the last two digits, for instance, for $b=101$ fractal the reading should be the following:   $B^*(s=1)=0.38815(05)\equiv 0.38815\pm0.00005$. The values for $b=\infty$ are obtained by  linear extrapolation of MCRG values.}\label{tabela1}
\begin{ruledtabular}
\begin{tabular}{@{}lllllll} 
$b$ & $B^*(s=1)$ & $B^*(s=0.9)$&  $B^*(s=0.7)$&$B^*(s=0.5)$ & $B^*(s=0.3)$ & $B^*(s=0.1)$\\
\hline
3 & 0.70711& 0.75595 & 0.85923&0.94815&0.99212 &0.99990\cr
5 & 0.59051& 0.63304& 0.73677&0.86443&0.97258 &0.99965\cr
7 & 0.53352& 0.57132 &0.66544 &0.79312&0.94262 &0.99924\cr
9 &0.50029& 0.53516 & 0.62208 &0.74257&0.90676 &0.99866\cr
11 & 0.47863(23) &0.51141(24) & 0.59352(26)& 0.70754(27)&0.87251(13)&0.99785(05)\cr
13 & 0.46319(14) &0.49449(20) & 0.57335(22)& 0.68289(24)&0.84351(25)&0.99679(07)\cr
15 & 0.45191(13) &0.48212(18) & 0.55857(20)& 0.66399(21)&0.82073(22)&0.99525(08)\cr
17 & 0.44321(11) &0.47307(17) & 0.54733(18)& 0.64984(19)&0.80218(20) &0.99315(09)\cr
21 & 0.43065(10) &0.45927(06) & 0.53091(07)& 0.62963(16)&0.77512(17)&0.98700(10)\cr
25 & 0.42207(08) &0.45005(12) & 0.51956(06)& 0.61549(14)&0.75663(14)&0.97767(11)\cr
31 & 0.41323(06) &0.44057(10) & 0.50811(11)& 0.60115(15)&0.73802(12)&0.96076(10)\cr
35 & 0.40913(06) &0.43611(09) & 0.50276(10)& 0.59441(11)&0.72923(11)&0.95010(10)\cr
41 & 0.40420(04) &0.43086(06) & 0.49664(04)& 0.58698(10)&0.71913(10)&0.93637(08)\cr
51 & 0.39893(07) &0.42498(07) & 0.48969(03)& 0.57825(15)&0.70797(09)&0.91977(07)\cr
61 & 0.39524(06) &0.42112(06) & 0.48493(07)& 0.57259(07)&0.70032(08)&0.90832(08)\cr
81 & 0.39081(05) &0.41638(05) & 0.47933(03)& 0.56545(07)&0.69079(11)&0.89413(05)\cr
101 & 0.38815(05) &0.41344(06)& 0.47601(11)& 0.56137(06)&0.68541(06)&0.88578(08)\cr
121 & 0.38649(04) &0.41167(07)& 0.47373(09)& 0.55862(05)&0.68184(05)&0.88007(04)\cr
151 & 0.38479(02) &0.40989(09)& 0.47163(04)& 0.55592(08)&0.67825(05)&0.87459(11)\cr
171 & 0.38399(03) &0.40894(06)& 0.47054(06)& 0.55464(04)&0.67668(04)&0.87217(07)\cr
201 & 0.38316(06) &0.40803(02)& 0.46951(11)& 0.55321(09)&0.67480(04)&0.86939(09)\cr
\vdots\cr
$\infty$& 0.37915(40)&0.40189(12)&0.46217(15)&0.54424(15)&0.66287(22)&0.85186(20)\cr
\end{tabular}
\end{ruledtabular}
\end{table*}
%

%
\begin{table*}
\caption{ The exact ($3\le b\le 9$) and
MCRG ($11\le b\le 201$) results for the critical exponents
$\nu$ obtained in this work, for the PF family of fractals.  Each MCRG entry of the Table has been
obtained by performing at least $10^5$ MC simulations. Numbers in the brackets correspond to the errors of the last two digits, determined by the MC simulation statistics (for example, for   $b=11$ and $s=0.9$, we have  $\nu=0.77239(27)\equiv 0.77239\pm 0.00027$).}\label{tabela2}
\begin{ruledtabular}
\begin{tabular}{@{}lllllll}
$b$ & $\nu(s=1)$& $\nu(s=0.9)$ & $\nu(s=0.7)$& $\nu(s=0.5)$ & $\nu(s=0.3)$ & $\nu(s=0.1)$ \\
\hline
3 & 0.79248& 0.81384 &0.87230 & 0.94400&0.99061 &0.99988\cr
5 & 0.78996& 0.79864&0.82629 &0.88266& 0.96922&0.99959\cr
7 & 0.78111& 0.78666 &0.80342 &0.83910&0.93233 &0.99906\cr
9 &0.77464& 0.77886 &0.79108 & 0.81500&0.88946 &0.99813\cr
11  &0.76959(27)&0.77239(27)&0.78194(26)&0.80127(27)&0.85478(15)&0.99659(03)\cr
13  &0.76494(18)&0.76923(25)&0.77739(25)&0.79250(25)&0.83325(29)&0.99415(11)\cr
15  &0.76232(17)&0.76571(24)&0.77307(24)&0.78678(24)&0.81825(26)&0.99022(14)\cr
17  &0.75976(17)&0.76226(23)&0.76962(23)&0.78205(22)&0.80885(24)&0.98385(18)\cr
21  &0.75522(16)&0.75754(10)&0.76411(09)&0.77496(21)&0.79717(21)&0.96152(27)\cr
25  &0.75199(15)&0.75406(07)&0.76001(09)&0.77000(19)&0.78993(20)&0.92615(31)\cr
31  &0.74822(12)&0.74954(20)&0.75559(19)&0.76515(18)&0.78315(18)&0.87448(29)\cr
35  &0.74530(14)&0.74773(19)&0.75278(18)&0.76276(17)&0.77889(17)&0.85207(25)\cr
41  &0.74332(08)&0.74487(06)&0.74966(08)&0.75837(16)&0.77439(16)&0.83262(21)\cr
51  &0.73969(17)&0.74148(17)&0.74614(07)&0.75381(15)&0.76970(15)&0.81726(17)\cr
61  &0.73643(17)&0.73930(16)&0.74352(05)&0.75051(14)&0.76579(14)&0.80992(15)\cr
81  &0.73314(15)&0.73527(14)&0.73935(06)&0.74640(19)&0.76050(12)&0.80041(13)\cr
101 &0.73103(11)&0.73280(04)&0.73610(14)&0.74313(12)&0.75689(12)&0.79482(12)\cr
121 &0.72865(13)&0.73030(22)&0.73459(28)&0.74071(11)&0.75405(11)&0.79097(11)\cr
151 &0.72771(07)&0.72840(06)&0.73219(11)&0.73848(10)&0.75102(10)&0.78618(08)\cr
171 &0.72631(12)&0.72752(05)&0.73094(10)&0.73830(10)&0.74850(09)&0.78338(10)\cr
201 &0.72486(12)&0.72633(05)&0.72958(12)&0.73583(04)&0.74755(09)&0.78077(10)\cr
\end{tabular}
\end{ruledtabular}
\end{table*}

To overcome the computational problem of learning exact
values of $P(N,N_b)$, for fractals with  $b\ge 11$,  we apply the
Monte Carlo renormalization group method (MCRG) \cite{d6}.  The
essence of the MCRG method  consists of treating $B'$, given by
(\ref{RGB}), as the grand canonical partition function that
accounts for all possible SAWs that traverse the fractal generator
at two fixed apexes. In this spirit, (\ref{RGB}) allows us to
write the following relation
\begin{equation}\label{b1}
{dB'\over dB}={B' \over B} \left< N(B)\right>\>
,
\end{equation}
where $\left< N(B)\right>$ is given by
\begin{equation}\label{b2}
\left< N(B,s)\right>={1\over B'}\sum_{N,N_b}NP(N,N_b) B^{N}s^{N_b}\>
,
\end{equation}
which can be considered as the average number of steps,
made with fugacity $B$ and stiffness $s$, by all possible SAWs that
cross the fractal generator. Then, from (\ref{RGB}) and (\ref{lamdai}), it follows
\begin{equation}\label{b3}
\lambda_\nu =\left< N(B^*,s)\right>.
\end{equation}

The last formula  enables us to calculate $\nu$ via the  MCRG
method, that is, without calculating explicitly the coefficients
$P(N,N_b)$. For a given fractal (with the scaling factor $b$) and
the SAW stiffness $s$, we begin with  determining  the
critical fugacity $B^*$.  To this end, we start the Monte  Carlo
(MC) simulation with an initial guess for the fugacity $B_0$  in
the  region $0<B_0 <1$.  Here $B_0$  can be interpreted as the
probability of making  the  next step along the same  direction
from the  vertex  that  the walker has reached, while $sB_0$ is
the probability to make the  next step by changing the step
direction. We assume that the walker starts his path at one
terminus (vertex) and tries to reach the other terminus of the
generator. In a case that the walker does not succeed to pass
through the generator, the corresponding path is not taken into
account. We repeat this MC simulation $L$ times, for the same set
$B_0$ and $s$. Thus, we find how many times the walker has passed
through the generator,
  and by dividing the corresponding number
by $L$ we get the value of the function (\ref{RGB}), denoted here by $B'(B_0,s)$.
In this way we get the value  of the sum (\ref{RGB}) without
specifying the set $P(N,N_b)$.  Then, for a fixed $s$, the  next  values $B_n$
$(n\ge 1)$, at which the MC simulation should be performed,
can  be found by using the ``homing'' procedure \cite{redner}, which
can be closed at the stage when the  difference  $B_n
-B_{n-1}$ becomes less than  the statistical  uncertainty
associated  with $B_{n-1}$.  Consequently, $B^*$  can be
identified with the last  value $B_n$ found in this way.
Performing the MC simulation at the values   $B^*$ and $s$, we can
record all possible  SAWs that traverse the fractal
generator. Then, knowing such a set of walks, we can
represent the average value of the length of a walk (that
traverse the generator) via the corresponding average
number of steps $\langle N(B^*,s)\rangle$, and,  accordingly, we
can learn  the value of the $\nu$
through the formulas  (\ref{b3}) and  (\ref{ni}).
In Tables \ref{tabela1} and \ref{tabela2}, we present our MCRG results for $B^*$ and $\nu$ respectively,  for the chosen set of $s$ values, for the PF fractal lattices with $11\le b\le 201$.

To calculate the critical exponent $\gamma$ we need to find the singular behavior of the generating  function $G(x,s)$.
The structure of the expression  (\ref{generatrisa}) shows
that the asymptotic behavior of $G(x,s)$, in the
vicinity of the critical fugacity $x_c(s)$, depends on
the corresponding behavior of the restricted partition
functions (\ref{eq:RGA})--(\ref{eq:RGC}). Assuming that
the singular behavior of (\ref{generatrisa}) is of the
form (\ref{g-sing}), it can be shown \cite{zivic3} that the critical
exponent $\gamma$ should be given by
\begin{equation}\label{gama}
\gamma=2{\ln (\lambda_\gamma/b)\over\ln \lambda_\nu}\>,
\end{equation}
 where $\lambda_\gamma$ is the RG eigenvalue
\begin{equation}\label{lamdag}
\lambda_\gamma=a(B^*,s)\> ,
\end{equation}
of the polynomial $a(B^{(n-1)},s)$ defined by (\ref{eq:RGA}), with
$B^*$ being the fixed point value of (\ref{RGB}). Therefore, it
remains either to determine exactly an
explicit expression for the polynomial
$a\left(B^{(n-1)},s\right)$, or somehow to surpass this step and to
evaluate  only the single needed value $a(B^*,s)$. We
have been able to determine the exact form of the requisite
polynomial for the PF fractals with $3 \le b \le 9$, while
for $b\ge11$ we have applied the MCRG to evaluate $a(B^*,s)$.
%
%
\begin{table*}
\caption{ The exact ($3\le b\le 9$) and
MCRG ($11\le b\le 201$) results for the critical exponents
$\gamma$ obtained in this work, for the PF family of fractals.  Each MCRG entry of the Table has been
obtained by performing at least $5\cdot10^5$ MC simulations.  The numbers in the brackets represent the error bars related to the  last two digits, for example for $b=201$ and $s=0.9$, we have  $\gamma=2.216(13)\equiv 2.216\pm0.013$.}\label{tabela3}
\begin{ruledtabular}
\begin{tabular}{@{}lllllll}
$b$ & $\gamma(s=1)$ & $\gamma(s=0.9)$ & $\gamma(s=0.7)$& $\gamma(s=0.5)$  & $\gamma(s=0.3)$&$\gamma(s=0.1)$ \\
\hline
3 & 1.6840& 1.6796 & 1.6056 & 1.3368& 0.8189 &0.2524\cr
5 & 1.7423& 1.7406& 1.7236 & 1.6250& 1.1654 &0.3363\cr
7 & 1.7614& 1.7605 & 1.7552 & 1.7247& 1.4586 &0.4302\cr
9 & 1.7807& 1.7795 & 1.7748 & 1.7596& 1.6335 &0.5340\cr
11  &1.8048(32) &1.7987(31) &1.7908(28) &1.7753(25) &1.7194(17) &0.6498(07)\cr
13  &1.8158(32) &1.8136(33) &1.8095(30) &1.8020(26) &1.7573(21) &0.7746(08)\cr
15  &1.8395(25) &1.8251(35) &1.8267(32) &1.8138(28) &1.7827(22) &0.9082(09)\cr
17  &1.8595(27) &1.8590(36) &1.8484(33) &1.8281(29) &1.7981(23) &1.0491(11)\cr
21  &1.8944(29) &1.8834(37) &1.8848(33) &1.8760(31) &1.8222(25) &1.3261(13)\cr
25  &1.9244(42) &1.9106(40) &1.9101(36) &1.8932(34) &1.8452(26) &1.5333(16)\cr
31  &1.9549(34) &1.9538(47) &1.9291(43) &1.9281(37) &1.8839(29) &1.6914(17)\cr
35  &1.9810(50) &1.9826(50) &1.9526(69) &1.9398(35) &1.9033(30) &1.7425(17)\cr
41  &1.9842(53) &1.9921(52) &1.9846(46) &1.9763(43) &1.9305(30) &1.7778(18)\cr
51  &2.0398(62) &2.0366(61) &2.0325(53) &1.9991(48) &1.9779(27) &1.8202(19)\cr
61  &2.0744(67) &2.0420(67) &2.0428(55) &2.0323(52) &1.9899(40) &1.8418(21)\cr
81  &2.0912(60) &2.0903(81) &2.0773(49) &2.0819(78) &2.0124(48) &1.8961(24)\cr
101 &2.124(17)  &2.1316(95) &2.1196(77) &2.1233(73) &2.0655(54) &1.9328(27)\cr
121 &2.172(11)  &2.158(12)  &2.145(11)  &2.1429(78) &2.1040(60) &1.9577(30)\cr
151 &2.175(13)  &2.182(13)  &2.182(10)  &2.174(10)  &2.1356(65) &1.9782(34)\cr
171 &2.191(14)  &2.198(15)  &2.192(13)  &2.1846(94) &2.1484(79) &2.0219(35)\cr
201 &2.2154(89) &2.216(13)  &2.2023(85) &2.219(14)  &2.1598(86) &2.0810(42)\cr
\end{tabular}
\end{ruledtabular}
\end{table*}
%

In order to learn an explicit expression of the polynomial
$a(B^{(r-1)},s)$, we note that its form, due to the underlying
self-similarity, should not depend on $r$, and, for this
reason, in what follows we assume $r=1$. Then, one can
verify the following expression
\begin{equation}\label{polinoma}
a(B,s)=\sum_{N,N_b} Q(N,N_b)B^Ns^{N_b}\>,
\end{equation}
where $Q(N,N_b)$ is the number of all SAWs of $N$ steps, with $N_b$ bends, that
start at any bond within the generator ($r=1$) and leave
it at a fixed exit.  By enumeration of all relevant walks,
the coefficients $Q(N,N_b)$ can be evaluated exactly up to
 $b=9$.
For $b=3$ fractal, the polynomial (\ref{polinoma}) is of the form
\begin{eqnarray}
a(B,s)&=&1+B+2Bs+B^2+2B^2s+2B^2s^2+2B^{3}s^2\nonumber\\
 &+&2B^{3}s^3+
4B^{4}s^{3}+4B^{4}s^{4}+2B^{5}s^{4}+2B^{6}s^{5} \>,\nonumber\\
 &&\label{Q3}
 \end{eqnarray}
for $b=5$ it is given in the Appendix, while for $b = 7$ and 9
they are given in the supplementary EPAPS Document \cite{supp}.
Using this  information, together with (\ref{lamdag}), (\ref{gama}), and previously found $B^*$ and $\lambda _\nu$, we have obtained the desired exact
values of $\gamma$ (see Table~\ref{tabela3}).

For a sequence of $b>9$, the exact determination of the
polynomial (\ref{polinoma}), that is, knowledge of the coefficients
$Q(N,N_b)$, can be hardly reached using the present-day
computers. However, to calculate $\lambda_\gamma$ one does not
need a complete knowledge of polynomial $a(B,s)$, but only its values at the fixed point (see Eq. (\ref{lamdag})). However,  the polynomial that appears in (\ref{eq:RGA}) can be
conceived as grand partition function of an appropriate
ensemble, and consequently, within the MCRG method,
the requisite value of the polynomial can be determined
directly \cite{zivic3}. Owing to the fact that we can obtain $\lambda_\gamma=a(B^*,s)$
through the MC simulations, and, knowing $\lambda_\nu$ from
the preceding calculation of $\nu$, we can apply (\ref{gama}) to calculate $\gamma$. In Table~\ref{tabela3} we present our MCRG
results of $\gamma$ for $11\le b\le201$, for the chosen set of  stiffness parameter values ($s=1$, 0.9, 0.7, 0.5, 0.3 and 0.1).

\section{Discussion and summary}
\label{treca}

We have studied critical properties of semiflexible polymer chains  on the
infinite family of the PF fractals whose each member has
the fractal dimension $d_f$ equal to the Euclidean value 2.
In particular, we have calculated the  critical
exponents $\nu$ and $\gamma$ via an exact RG  (for $3\le
b\le 9$) and via the MCRG approach (up to $b=201$). Specific results
for the  critical exponents have been presented in
Tables~\ref{tabela2} and \ref{tabela3}.
\begin{figure*}
\centerline{\includegraphics[width=6.5in]{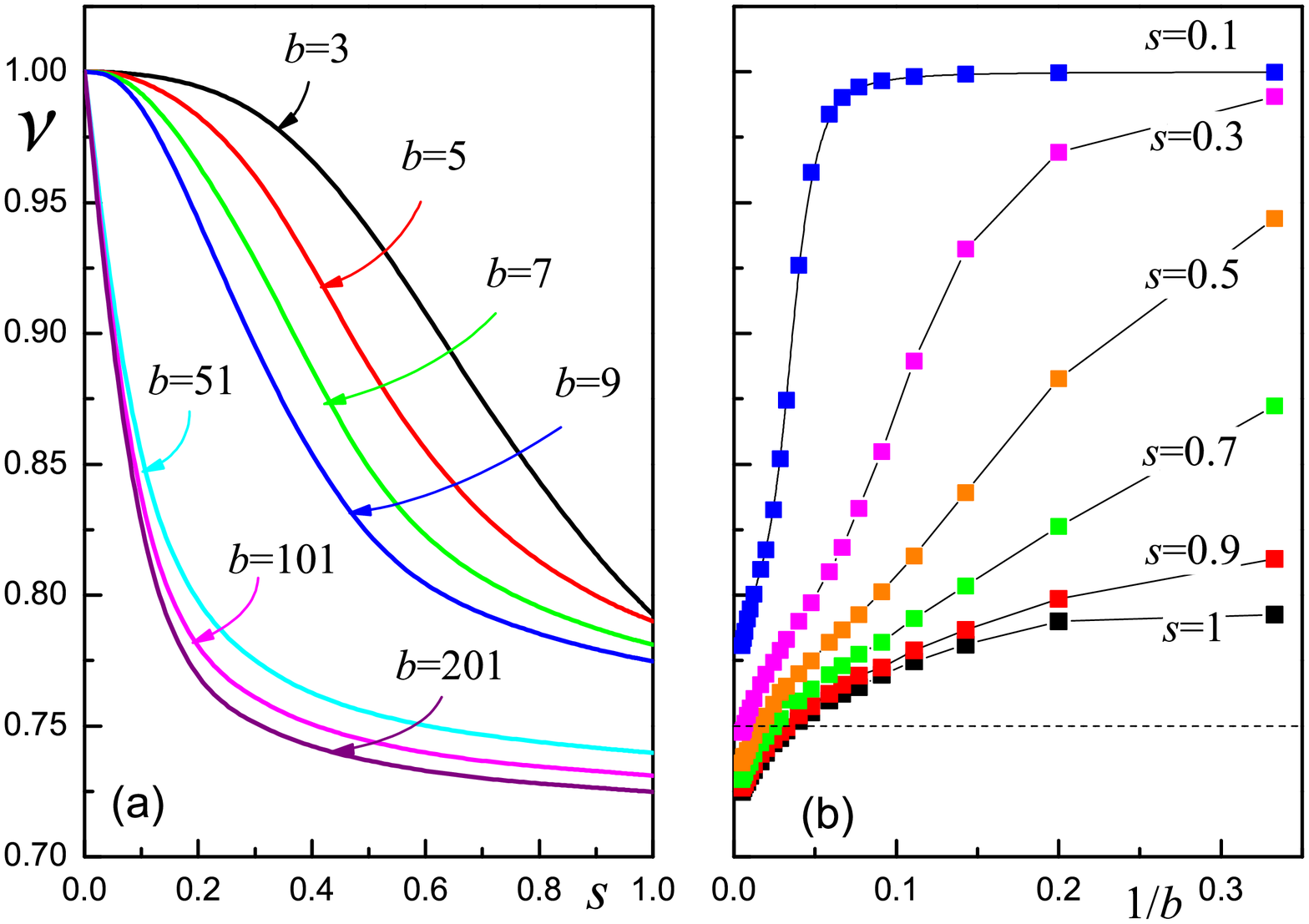}}
\caption{ \label{fig:ni} (Color online)  Plot of the end-to-end distance critical exponent $\nu$ of semiflexible polymer chain on PF fractals: (a) as a function of stiffness parameter $s$, for various fractal scaling parameter $b$, and (b) as a function of $1/b$, for various values of stiffness parameter $s$ (the horizontal
 dashed line represents the Euclidean value $\nu =3/4$ for flexible polymers, $s=1$, whereas thin solid lines  serve only as guides to the eye).}
\end{figure*}

In order to analyze the obtained results, in Fig.~\ref{fig:ni}(a) we
have plotted $\nu$  as a function of stiffness parameter $s$, for
several values of fractal scaling parameter $b$. One can see that
for each $b$, exponent $\nu$ monotonically decreases from the
value $\nu=1$, for $s=0$, corresponding to rigid rod, to the value
$\nu_{SAW}(b)$, for $s=1$, corresponding to the flexible polymer
chain \cite{z12,zivic3}.  This, indeed, implies that for finite
$b$, the mean end-to-end distance for semiflexible polymers
increases with its rigidity, and is always between its values for
the flexible chain and the rigid rod. In the same figure one can
also observe that when $b$ increases the curves $\nu(s)$ become
increasingly sharper so that their limit looks to  be $\nu=1$, at
$s=0$, whereas $\nu\approx \text{const.}$, for $0<s\le 1$. This
observation may imply that for very large $b$ (beyond $b=201$)
the critical exponent $\nu$ becomes independent of $s$. Here, one should note that it is
believed that critical exponent $\nu$ is universal for
semiflexible SAWs on Euclidean lattices, that is, it does not
depend on $s$ \cite{giacometti}. This expectation is based on
universality arguments, and it was exactly demonstrated for
directed semiflexible SAWs \cite{privmanKnjiga}. The same
conclusion was also exactly derived for semiflexible SAWs on the
Havlin--Ben-Avraham and 3-simplex fractal lattices
\cite{giacometti}. However, as it was pointed out in
\cite{giacometti}, in contrast to the case of homogeneous
lattices, where rigidity only increases the persistence length,
but does not affect neither the scaling law governing the critical
behavior of the the mean end-to-end distance, nor the value of the
critical exponent $\nu$ of SAWs, one might expect that presence of
disorder in nonhomogeneous lattices, combined with the stiffness,
in some cases can constrain the persistence length, and
consequently induces dependence of $\nu$ on $s$. This was
explicitly confirmed in the same paper, by exact calculation of
the critical exponent $\nu$ for branching Koch curve, which turned
out to be continuously decreasing function of $s$, similar to
functions  depicted in  Fig.~\ref{fig:ni}(a).
Apparently, the established
dependence of $\nu$ on $s$ for PF fractals  with smaller values of $b$ shows that considerable lattice disorder affects significantly   the values of $\nu$, while the dependence of $\nu$ on $s$ gradually disappears for PF fractals with smaller disorder (appearing for larger  $b$). These facts
confirm the assumption \cite{giacometti}  that lattice disorder, combined with the polymer stiffness, has a predominant impact on the
critical behavior of semiflexible polymers.

In Fig.~\ref{fig:ni}(b) data for $\nu$ as a function of $1/b$ are
depicted, for various values of $s$. It appears that for each
considered value of stiffness $s$ in the range $0<s\le 1$, the
critical exponent $\nu$ is monotonic function of the scaling
parameter $b$,  in the region of $b$ studied.  It can be also seen that
for large fixed $b$, the differences between the values of
$\nu(s)$ (for various $s$) decrease when $b$ increases, which
brings us to the question of the behavior of $\nu$ in the
fractal--to--Euclidean crossover, when $b\to\infty$. Concept of the fractal-to-Euclidean crossover is often used in order to study
if and how various physical properties change when inhomogeneous 
lattices approach homogeneous (translationally invariant
Euclidean) lattice. By tuning some conveniently chosen parameter
of the fractal lattice (such as scaling parameter $b$ in the case
of PF fractals), properties of the corresponding Euclidean lattice
(square lattice in this case) can be gradually approached. Studies
of  the flexible SAW models on Sierpinski gasket family of fractals
\cite{d3,d6,p1,p2,p3,d7,d8}, as well as on PF fractals \cite{zivic3},
revealed that crossover behavior of critical exponents can be
rather subtle in the sense that not all critical exponents tend to
their Euclidean values, and even when they do so it can be
accomplished in quite unexpected manner. For instance, according
to the finite--size scaling arguments, when $b\to\infty$ exponent
$\nu$ of flexible polymers ($s=1$) on PF fractals approaches the
Euclidean value 3/4 from below \cite{zivic3}, which together with
the fact that $\nu$ is monotonically decreasing function for $b$
up to $201$, means that for some value of $b$ larger than 201
there should exist a minimum. On the other hand, for $s=0$
exponent $\nu$ is equal to 1 for each $b$.  For $0<s\leq 1$
apparent trend of the curves presented in Fig.~\ref{fig:ni}(b)
suggests that limiting value of $\nu$, when $b\to\infty$, does not
depend on particular value of $s$, and following the behavior of $\nu$ for flexible polymers, it should be equal to the Euclidean value $3/4$. However,  we would not like to draw here such a definite conclusion without
additional investigations.
\begin{figure*}
\centerline{\includegraphics[width=6.5in]{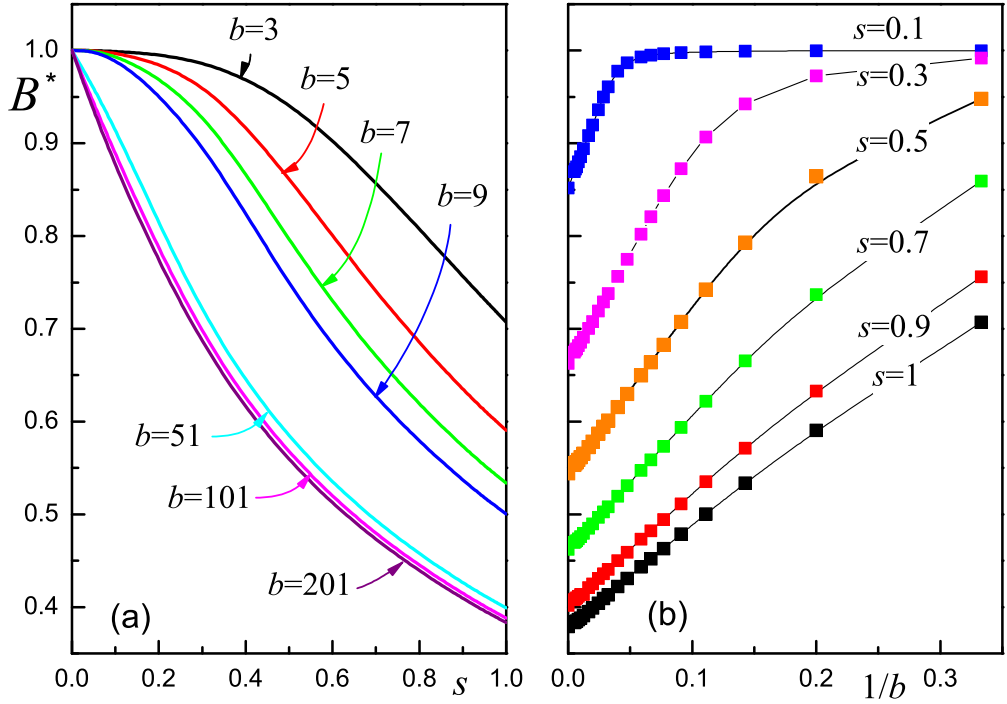}}
\caption{ \label{fp} (Color online)  Data for the fixed point values $B^*$ (the
reciprocal connectivity)  of semiflexible polymer chain plotted as: (a) function of stiffness $s$, for various values of the fractal parameter $b$, and (b) function of  $1/b$, for various values of $s$. }
\end{figure*}

\begin{figure*}
\centerline{\includegraphics[width=6.5in]{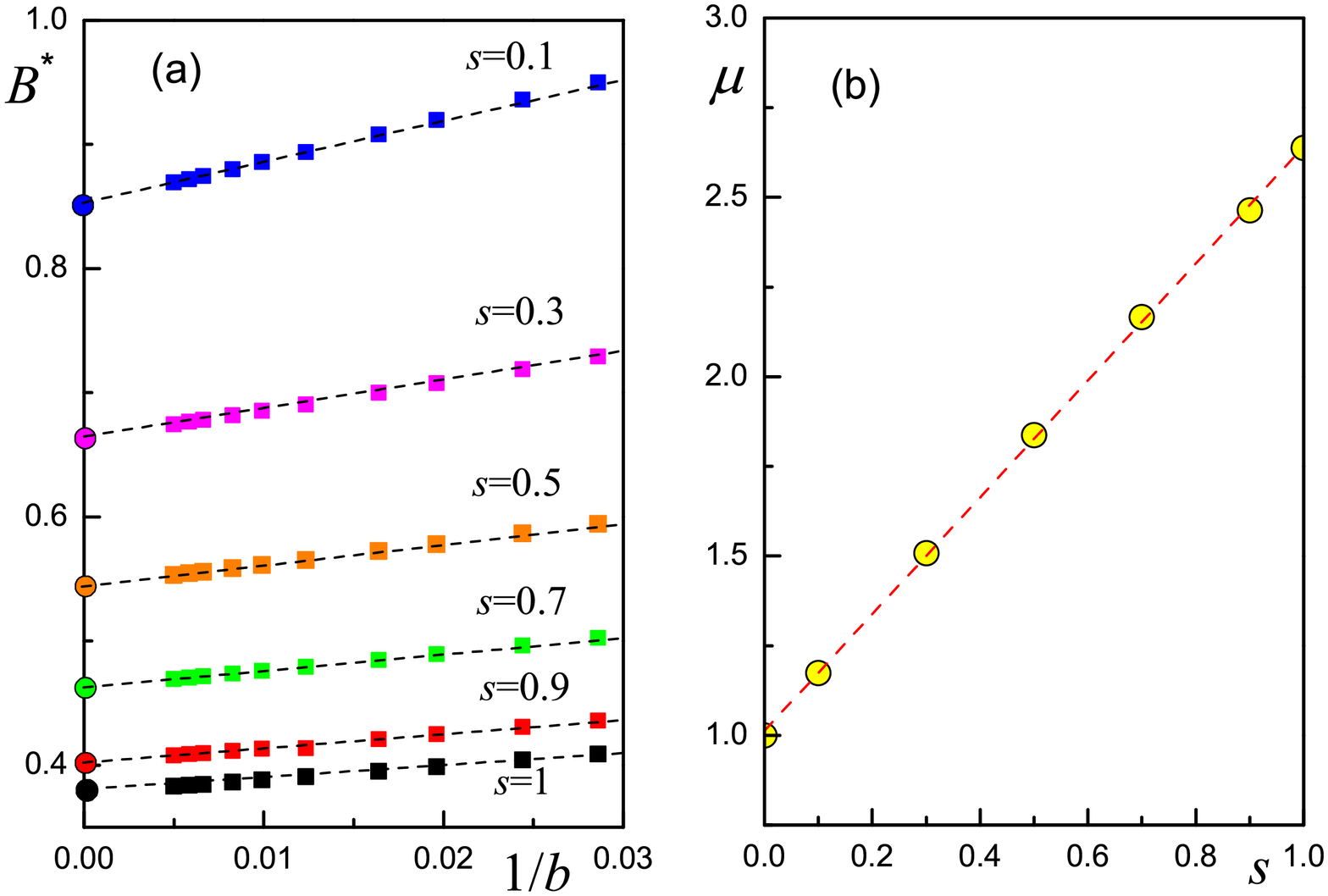}}
\caption{ \label{fig:fit} (Color online)  (a) Linear fitting of the data (full squares) for the fixed point $B^*$ (see Table~\ref{tabela1}) as function of $1/b$ for large values of $b$, and various $s$. Circles correspond to the extrapolated limiting values of $B^*(s,b\to\infty)$, and their precise values are given in the last row of Table~\ref{tabela1}. (b) Plot of the connectivity constant $\mu(s)=1/B^*(s,b\to\infty)$. Dashed line is the linear fit of the data presented by circles, which correspond to $\mu(s)$ for $s=0$, 0.1, 0.3, 0.5, 0.7, 0.9 and $s=1$. }
\end{figure*}

Continuing the comparison of the criticality of flexible and semiflexible  SAWs on the PF fractals, in Fig.~\ref{fp} we have depicted
the data from Table~\ref{tabela1} for the critical fixed points $B^*$.
On the left-hand side of this figure one can notice that $B^*$, which is equal to the reciprocal of the connectivity constant $\mu$, is monotonically decreasing function of $s$, for each $b$ considered. This has been expected, since $\mu$ has the physical meaning of the average number of steps available to the walker having already completed a large number of steps, so that larger flexibility of the polymer chain implies larger $\mu$, and consequently $\mu(s=0)<\mu(0<s<1)<\mu(s=1)$.
In Fig.~\ref{fp}(b) one can also observe that for fixed $s$, the fixed point $B^*$ decreases with $b$. Furthermore, $B^*$ becomes almost linear function of $1/b$ for large $b$, which allows us to estimate the limiting values of
$B^*$  for $b\to\infty$. The obtained  asymptotic values are given at the end of Table~\ref{tabela1}.
The value $B^*(s=1)=0.37915\pm0.00040$, should be compared with the Euclidean value
0.3790523(3) for the square lattice, obtained in \cite{guttmann}.
As one can see, the agreement is very good, and we may say that
for flexible polymers the values $B^*(b)$, in the limit $b\to\infty$, converge
to the Euclidean value.  Similarly, we expect that, for given $s< 1$, the
values $B^*(b)$ of semiflexible polymers also converge to the
corresponding $d=2$ Euclidean values (which are functions of $s$).
In Fig.~\ref{fig:fit}(a) we have plotted the lines obtained by linear
fitting of the large $b$ data for $B^*(s)$ for $s=0.1$, 0.3, 0.5, 0.7, 0.9 and $s=1$.
In the present situation, estimated limiting values of the connectivity constants
$\mu(s)=1/B^*(s,b\to\infty)$ are depicted in Fig.~\ref{fig:fit}(b), as
function of $s$. It seems that $\mu(s)$ is linear function of $s$, implying
that connectivity constant for semiflexible SAWs on square lattice could be a
linear function of the stiffness parameter $s$. Such expectation is also in
accord with the exact results obtained for directed semiflexible SAWs
\cite{privmanKnjiga}.
\begin{figure*}
\centerline{\includegraphics[width=6.5in]{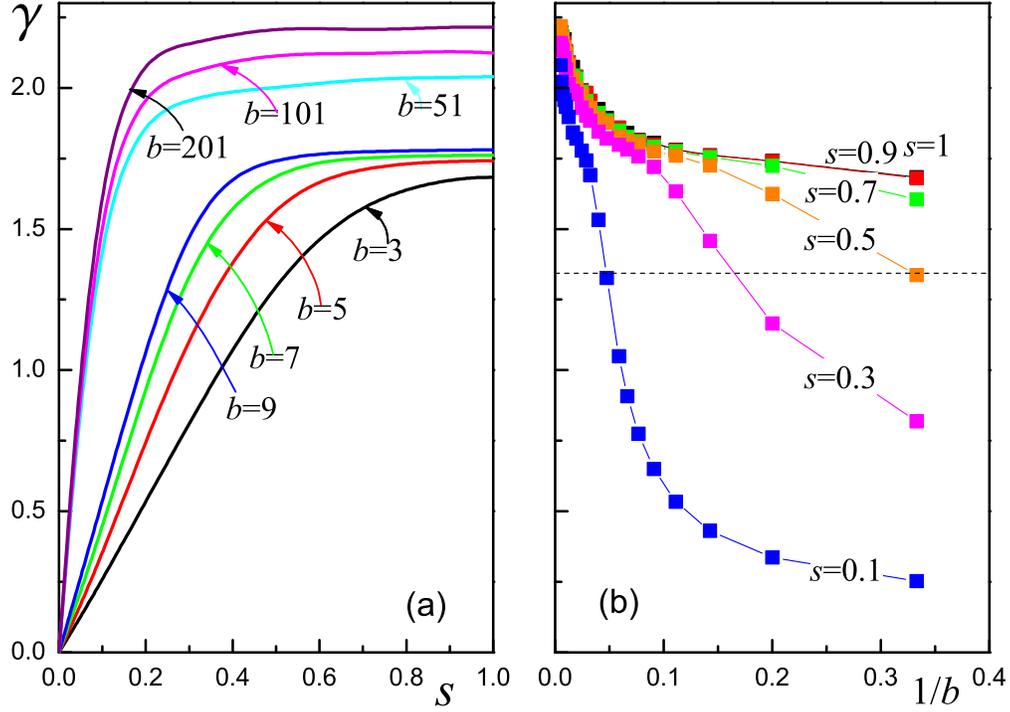}}
\caption{ \label{fig:gamma} (Color online)  Data for the SAW critical exponent $\gamma$  presented as: (a)  function of the stiffness parameter $s$, for various values of the fractal enumerator $b$, and (b)  function of $1/b$, for  $s=1$, 0.9, 0.7, 0.5, 0.3, and 0.1 (the horizontal dashed line represents the two--dimensional Euclidean value $\gamma =43/32$ for flexible polymers $s=1$).}
\end{figure*}

To make our analysis of  semiflexible  SAWs on PF fractals complete, in Fig.~\ref{fig:gamma} we present the data found for the critical exponent $\gamma$. 
As it was explained in Sec.~\ref{druga}, exponent $\gamma$ is given by
(\ref{gama}): $\gamma=2\ln(\lambda_\gamma/b)/\ln\lambda_\nu$, where
eigenvalues $\lambda_\nu$  and $\lambda_\gamma$ (given by (\ref{lamdai}) and
(\ref{lamdag}), respectively) are evaluated at the fixed point $B=B^*$ of the
RG equation (\ref{eq:RGB}).
For $b=3$ the fixed point $B^*$ is given by (\ref{ng3a}), and since dependence of $\lambda_\nu$ and $\lambda_\gamma$ on $B$ is known by exact means, the curve $\gamma(b=3,s)$ in Fig.~\ref{fig:gamma}(a) was plotted according to the closed-form exact formula. For $b=5$, 7 and 9, RG equation (\ref{eq:RGB}) was also found explicitly, but its fixed point can be calculated only numerically in these cases. Nevertheless, this task can be done for large number of $s$ values, and putting fixed points $B^*(s)$ calculated accordingly, into the exact expressions found for $\lambda_\nu$ and $\lambda_\gamma$, one obtains the corresponding values for $\gamma$, and, consequently, curves for $b=5$, 7 and 9 in Fig.~\ref{fig:gamma}(a). For larger values of $b$, depicted $\gamma$ curves were obtained by interpolating the data found by MCRG approach for $s=0.1$, 0.3, 0.5, 0.7, 0.9 and $s=1$ (Table~\ref{tabela3}), and generalizing the fact $\gamma(b,s=0)=0$, exactly found for smaller values of $b$, to all $b$ values. One can see that for each $b$, the critical exponent $\gamma$ is monotonically increasing function of the stiffness parameter $s$. The dependence of $\gamma$ on $s$ is in accord with the discussed non-universality of $\nu$, and also with the results obtained for $\gamma$ of SAWs on the branching Koch curve and Havlin--Ben-Avraham fractal~\cite{giacometti}. However, one should note here that while for the Koch curve both $\nu$ and $\gamma$ depend on $s$, in the case of Havlin--Ben-Avraham fractal only $\gamma$ is non-universal ($\nu=1$ for all values of $s$). Besides, in \cite{giacometti} it was shown that neither $\nu$ nor $\gamma$ depend on $s$ for SAWs on the 3-simplex fractal. The observed different behavior of exponents $\nu$ and $\gamma$ on various fractals is an intriguing fact and imposes the question of the universality of $\gamma$ for semiflexible SAWs on homogeneous lattices. One might try to draw a helpful conclusion by looking at the large $b$ behavior of the functions $\gamma(s)$, plotted in Fig.~\ref{fig:gamma}(a). One can see that as $b$ grows the curve $\gamma(s)$ becomes sharper, so that for $b=201$ it is almost constant in the large part of the region $0<s\leq 1$, whereas  in the vicinity of $s=0$ it rapidly drops to the value $\gamma=0$, at $s=0$. Therefore, it may be concluded  that for $b\gg 1$ and $0<s\leq 1$ exponent $\gamma$ becomes independent of $s$. Furthermore, in Fig.~\ref{fig:gamma}(b),  we perceive that for each studied $s$, the critical exponents $\gamma$ monotonically increase  with $b$, and for $b=201$ acquire  almost the same value $\gamma\approx 2.2$. These observations may imply that $\gamma$ for semiflexible SAWs on homogeneous lattices is universal. However, it is known that critical exponent $\gamma$ for flexible polymers ($s=1$) on the  two-dimensional Euclidean lattices is equal to $\gamma=43/32$, which is far from the apparent limiting value 2.2 (suggested by the plots in Fig.~\ref{fig:gamma}(b), when $1/b\to 0$), implying that $\gamma$ for PF fractals does not tend to its Euclidean value for large $b$. This may seem odd, but it fits quite well into the peculiar picture which have emerged during the last two decades  for the fractal-to-Euclidean crossover behavior of the exponent $\gamma$ of flexible SAWs on PF~\cite{zivic3} and on Sierpinski gasket (SG) family of fractals~\cite{d3,d7,d8}, as well as for some  models of directed  SAWs on SG fractals~\cite{p3}. For instance, using finite-size scaling arguments, Dhar~\cite{d3} concluded that $\gamma$ for SAWs on SG fractals at the fractal-to-Euclidean crossover approaches the non-Euclidean value 133/32. In a similar manner, for PF fractals it was also demonstrated \cite{zivic3}
that in the limit $b\to\infty$ exponent $\gamma$ tends to $103/32$, which is again the non-Euclidean value.
In addition, numerical analysis of the large set of exact values of $\gamma$ obtained for the piece-wise directed SAWs on SG fractals, as well as an exact asymptotic analysis~\cite{p3}, also showed that the limiting value of $\gamma$ differs from the corresponding Euclidean value. In all these cases the established crossover behavior could not have been predicted only on the basis of $\gamma$ values obtained for relatively small $b$ (up to $b=201$, for instance).  On these grounds we may infer that in the crossover region, when $b\to\infty$, critical exponent $\gamma$ does not depend on the stiffness $s$, and approaches the non-Euclidean value.

In conclusion, we may say that family of plane-filling fractals
proved to be useful for investigation of the effects of the rigidity on the criticality  of SAWs on nonhomogeneous lattices. It is amenable to applying
exact and Monte Carlo renormalization group study, which we performed on large number of its members. The obtained
results show that the critical behavior of semiflexible SAWs is not universal, in a sense that critical end-to-end
exponents $\nu$, as well as entropic exponents $\gamma$, continuously vary with the stiffness parameter $s$.  Such
non-universality does not occur on regular lattices, but it was found for SAWs on the branching Koch curve,
suggesting that polymer behavior in realistic disordered environment might be more affected by its stiffness than it
was expected. Apart from the stiffness parameter $s$, critical exponents also depend on the fractal parameter $b$, but the trend of functions $\nu(s)$ and $\gamma(s)$ is similar for different $b$ values. This similarity becomes
more pronounced as $b$ grows and approaches the fractal-to-Euclidean crossover region ($b\to\infty$). Assuming
that critical exponents on regular lattices do not depend on $s$, it would be challenging  to reveal what
exactly happens with the exponents in the limit $b\to\infty$, which we would like to investigate  in the future.

\begin{acknowledgments}
{This paper has been done as a part of the
work within the project No. 141020B funded by the Serbian Ministry
of Science.}
\end{acknowledgments}

\appendix*
\section{}

Here we give the coefficients $Q(N,N_b)$ of the RG equation (\ref{polinoma})  for $b=5$ PF fractal:

 $Q(    0,    0)=         1 $, $Q(    1,    0)=         1 $, $Q(    1,    1)=         2 $, $Q(    2,    0)=         1 $, $Q(    2,    1)=         2 $, $Q(    2,    2)=         2 $, $Q(    3,    0)=         1 $, $Q(    3,    1)=         4 $, $Q(    3,    2)=         6 $, $Q(    3,    3)=         4 $, $Q(    4,    0)=         1 $, $Q(    4,    1)=         4 $, $Q(    4,    2)=        10 $, $Q(    4,    3)=        12 $, $Q(    4,    4)=         6 $, $Q(    5,    2)=         6 $, $Q(    5,    3)=        18 $, $Q(    5,    4)=        18 $, $Q(    5,    5)=         6 $, $Q(    6,    2)=         2 $, $Q(    6,    3)=        20 $, $Q(    6,    4)=        38 $, $Q(    6,    5)=        32 $, $Q(    6,    6)=        12 $, $Q(    7,    2)=         2 $, $Q(    7,    3)=         8 $, $Q(    7,    4)=        28 $, $Q(    7,    5)=        34 $, $Q(    7,    6)=        32 $, $Q(    7,    7)=        10 $, $Q(    8,    3)=         4 $, $Q(    8,    4)=        16 $, $Q(    8,    5)=        52 $, $Q(    8,    6)=        62 $, $Q(    8,    7)=        48 $, $Q(    8,    8)=        14 $, $Q(    9,    4)=        10 $, $Q(    9,    5)=        34 $, $Q(    9,    6)=        70 $, $Q(    9,    7)=        54 $, $Q(    9,    8)=        34 $, $Q(    9,    9)=         6 $, $Q(   10,    5)=        18 $, $Q(   10,    6)=        38 $, $Q(   10,    7)=        78 $, $Q(   10,    8)=        68 $, $Q(   10,    9)=        44 $, $Q(   10,   10)=         4 $, $Q(   11,    5)=         2 $, $Q(   11,    6)=        14 $, $Q(   11,    7)=        44 $, $Q(   11,    8)=        70 $, $Q(   11,    9)=        52 $, $Q(   11,   10)=        36 $, $Q(   11,   11)=         6 $, $Q(   12,    7)=        18 $, $Q(   12,    8)=        30 $, $Q(   12,    9)=        56 $, $Q(   12,   10)=        42 $, $Q(   12,   11)=        34 $, $Q(   12,   12)=         6 $, $Q(   13,    8)=         4 $, $Q(   13,    9)=        20 $, $Q(   13,   10)=        28 $, $Q(   13,   11)=        30 $, $Q(   13,   12)=        22 $, $Q(   13,   13)=         2 $, $Q(   14,   11)=        22 $, $Q(   14,   12)=        22 $, $Q(   14,   13)=        12 $, $Q(   15,   12)=         4 $, $Q(   15,   13)=         8$.


\end{document}